\documentclass[prd,aps,showpacs,epsf,floats,onecolumn]{revtex4}%
\usepackage{amssymb}
\usepackage{amsfonts}
\usepackage{amsmath}
\usepackage{graphicx}%
\providecommand{\U}[1]{\protect\rule{.1in}{.1in}}

\begin{document}
\title{\textbf{On the violation of Bell's inequality for all non-product quantum
states}}
\author{\textbf{Carlo Cafaro}$^{1}$, \textbf{Sean Alan Ali}$^{2}$, and \textbf{Adom
Giffin}$^{3}$}
\affiliation{$^{1}$SUNY Polytechnic Institute, 12203 Albany, New York, USA }
\affiliation{$^{2}$Albany College of Pharmacy and Health Sciences, 12208 Albany, New York, USA}
\affiliation{$^{3}$Clarkson University, 13699 Potsdam, New York, USA}

\begin{abstract}
We present an explicit reexamination of Gisin's 1991 original proof concerning
the violation of Bell's inequality for any pure entangled state of
two-particle systems. Given the relevance of Gisin's work, our analysis is
motivated by pedagogical reasons and allows the straightening of a few
mathematical points in the original proof that in no way change the physical
conclusions reached by Gisin.

\end{abstract}

\maketitle

\section{Introduction}

Relativity theory, quantum theory, and information theory are fundamental
blocks of theoretical physics \cite{peres2004}. The goal of theoretical
physics is to describe and, to a certain extent, understand natural phenomena.
Unfortunately, complex difficulties arise when one attempts to merge general
relativity theory and quantum theory. For example, in classical mechanics it
is often said that gravity is a purely geometric theory since the mass does
not appear in the usual Newtonian equation of a particle trajectory,%
\begin{equation}
m\frac{d^{2}\vec{x}}{dt^{2}}=-m\vec{\nabla}_{\vec{x}}\Phi_{\text{gravity}%
}\Leftrightarrow\frac{d^{2}\vec{x}}{dt^{2}}=\vec{g}\text{.}%
\end{equation}
This is a direct consequence of the equality of the gravitational and inertial
masses. In quantum mechanics, the situation is rather different. As a matter
of fact, the Schrodinger quantum-mechanical wave equation is given by
\cite{sakurai},%
\begin{equation}
\left[  -\frac{\hbar^{2}}{2m}\vec{\nabla}_{\vec{x}}^{2}+m\Phi_{\text{gravity}%
}\right]  \psi\left(  \vec{x}\text{, }t\right)  =i\hbar\frac{\partial
}{\partial t}\psi\left(  \vec{x}\text{, }t\right)  \text{.}%
\end{equation}
The mass $m$ no longer cancels and, instead, it appears in the combination
$\hbar/m$ (where $\hbar\overset{\text{def}}{=}h/2\pi$ and $h$ denotes the
Planck constant). Therefore, in an instance where $\hbar$ appears, $m$ is also
expected to appear \cite{colella1975}. It seems evident that there is the
possibility that such difficulties are not simply technical and mathematical,
but rather conceptual and fundamental. This viewpoint was recently presented
in \cite{brukner14}, where the idea was advanced that \emph{quantum causality}
might shed some light on foundational issues related to the general
relativity-quantum mechanics problem.

When describing natural phenomena at the quantum scale, the interaction
between the mechanical object under investigation and the observer (or,
observing equipment) is not negligible and cannot be predicted \cite{bohr35,
bohr37, bohr50}. This fact leads to the impossibility of unambiguously
distinguishing between the object and the measuring instruments. This, in
turn, is logically incompatible with the classical notion of causality; the
possibility of sharply distinguishing between the subject and the object is
essential to the ideal of causality. In his attempt to bring consistency in
science, Bohr proposed to replace the classical ideal of causality with a more
general viewpoint termed \emph{complementarity}. Roughly speaking, anyone can
understand that one cannot bow in front of somebody without showing one's back
to somebody else. This oversimplified statement is behind one of the most
revolutionary scientific concepts of the twentieth century, namely Bohr's
complementarity principle \cite{wheeler63}. This principle is a key feature of
quantum physics and represents the dichotomy between the corpuscular
(particle) and ondulatory (wave) nature of mechanical objects (matter and
light). Within this descriptive framework, particle and wave properties are
symbolized by well-defined position and momentum, respectively
\cite{vaccaro10}.

In 1935, using the complementarity principle, Bohr criticized the Einstein,
Podolsky, and Rosen conclusion according to which the quantum mechanical
description of physical reality given by wave functions was not complete based
on a line of reasoning relying on a thought experiment (gedankenexperiment)
\cite{epr}. In addition to criticizing the Einstein-Podolsky-Rosen
argumentation (EPR paradox), Bohr provided a different interpretation of the
concept of locality \cite{bohr35}. The solution to the EPR paradox is due to
John Bell's 1964 theorem \cite{bell64}. The notion of causality played a
key-role in the EPR paradox, Bohr's complementarity principle, and Bell's theorem.

In this article, we reexamine Gisin's 1991 original proof concerning the
violation of Bell's inequality for any pure entangled state of two-particle
systems \cite{gisin91}. Our investigation is motivated by didactic reasons and
permits to straighten a few mathematical points in the original proof that in
no way modify the physical content provided by Gisin's work.

\section{Background}

Einstein, Podolsky, and Rosen considered a composite quantum system consisting
of two distant particles, with an entangled wave function $\psi$ given by
\cite{epr},
\begin{equation}
\psi=\delta\left(  x_{1}-x_{2}-L\right)  \delta\left(  p_{1}+p_{2}\right)
\text{.}\label{wave}%
\end{equation}
The following background discussion follows very closely the presentation
presented by Asher Peres in Ref. \cite{peres95}. The quantity $\delta$ denotes
a normalizable function with an arbitrarily high and narrow peak. The quantity
$L$ is a large distance, much larger than the range of mutual interaction of
particles $1$ and $2$. The physical meaning of the wave function in Eq.
(\ref{wave}) is that the two particles have been prepared in such a way that
their relative distance is arbitrarily close to $L$, and their total momentum
is arbitrarily close to zero. Note that the operators $x_{1}-x_{2}$ and
$p_{1}+p_{2}$ commute. In the state $\psi$, one knows nothing about the
positions of the individual particles (we only know their distance from each
other); and one knows nothing of their individual momenta (one only knows the
total momentum). However, if one measures $x_{1}$, one shall be able to
predict with certainty the value of $x_{2}$, without having in any way
disturbed particle $2$. At this point, EPR argue that \textit{since at the
time of measurement the two systems no longer interact, no real change can
take place in the second system in consequence of anything that may be done to
the first system}. Therefore, $x_{2}$ corresponds to an element of physical
reality, as defined by EPR. On the other hand, if one prefers to perform a
measurement of $p_{1}$ rather than $x_{1}$, one shall then be able to predict
with certainty the value of $p_{2}$, again without having in any way disturbed
particle $2$. Therefore, by the same argument as above, $p_{2}$ also
corresponds to an element of reality. However, quantum mechanics precludes the
simultaneous assignment of precise values to both $x_{2}$ and $p_{2}$, since
these operators do not commute, and thus EPR are forced to conclude that the
quantum mechanical description of physical reality given by wave functions is
not complete. However, they prudently leave open the question of whether or
not a complete description exists. Reality according to EPR can be described
as follows: If, without in any way disturbing a system, one can predict with
certainty the value of a physical quantity, then there exists an element of
reality corresponding to this physical quantity.

The EPR article was not wrong, but it had been written too early
\cite{peres04}. The EPR argument did not take into account that the observer's
information was localized, like any other physical object. Information is not
just an abstract notion. It requires a physical carrier, and the latter is
approximately localized.

Let us recall that Einstein's locality principle asserts that events occurring
in a given spacetime region are independent of external parameters that may be
controlled, at the same moment, by agents located in distant spacetime
regions. In quantum mechanics, one has to accept that a measurement on what
seems to be a part of the system is to be considered as a measurement on the
whole system. If one persists with keeping Einstein's locality principle,
alternative theories incorporating such a principle lead to a testable
inequality (Bell's inequality, \cite{bell64}) relation among suitable
observables that are not in agreement with the predictions of quantum
mechanics. Several violations of Bell's inequality have been experimentally
verified \cite{peres95}. For this reason, despite the psychological
uncomfortable situation, quantum mechanics has prevailed over alternative
theories.\emph{ Dura lex sed lex}: this is the experimental verdict. Quantum
mechanics predictions are incompatible with Bell's inequality. There is an
experimentally verifiable difference between quantum mechanics and the
alternative theories satisfying Einstein's locality principle. It is somewhat
of ironic that Bell's theorem is the most profound discovery of science
because it is not obeyed by experimental facts \cite{stapp75}. We remark that
Bell's paper is not about quantum mechanics. Rather, it is a general proof,
independent of any specific physical theory, that there is an upper limit to
the correlation of distant events, if one solely assumes the validity of
Einstein's principle of local causes. Specifically, Bell showed that in a
theory in which parameters are added to determine the results of individual
measurements, without changing the statistical predictions, there must be a
mechanism whereby the setting of one measuring device can influence the
reading of another instrument, however remote. Moreover, the signal involved
must propagate instantaneously, so that such a theory could not be Lorenz
invariant. In particular, for any nonfactorable quantum state $\psi$, it is
possible to find pairs of observables whose correlations violate Bell's
inequality (see Eq. (\ref{bell})). This means that, for such a state, quantum
theory makes statistical predictions which are incompatible with the demand
that the outcomes of experiments performed at a given location in space be
independent of the arbitrary choice of other experiments that can be
performed, simultaneously, at distant locations (this apparently reasonable
demand, as stated earlier, is the principle of local causes, also called
Einstein locality). Bell's theorem implies that quantum mechanics is
incompatible with the view that physical observables possess pre-existing
values independent of the measurement context. A hidden variable theory which
would predict individual events must violate the canons of special relativity:
there would be no covariant distinction between cause and effect. The EPR
paradox is resolved in the way which Einstein would have liked least.

\section{Reexamination}

Here, we choose to present Bell's Theorem as presented by Gisin in
\cite{gisin91} (for a later and stronger derivation by Gisin and Peres, we
refer to \cite{gisin92}). In what follows, whenever helpful to the discussion,
we shall be using the conventional Dirac bra-ket notation where a wave vector
$\psi$ will be denoted as $\left\vert \psi\right\rangle $.

Gisin's Theorem can be stated as follows \cite{gisin91}:\ Let $\psi
\in\mathcal{H}_{1}\otimes\mathcal{H}_{2}$. If $\psi$ is entangled (i.e. $\psi$
is not a product), then $\psi$ violates Bell's inequality. In other words,
there are projectors $a$, $a^{\prime}$, $b$, $b^{\prime}$, such that%
\begin{equation}
\left\vert P\left(  a\text{, }b\right)  -P\left(  a\text{, }b^{\prime}\right)
\right\vert +P\left(  a^{\prime}\text{, }b\right)  +P\left(  a^{\prime}\text{,
}b^{\prime}\right)  >2\text{,} \label{bell}%
\end{equation}
where%
\begin{equation}
P\left(  a\text{, }b\right)  \overset{\text{def}}{=}\left\langle \left(
2a-1\right)  \otimes\left(  2b-1\right)  \right\rangle _{\psi}\text{.}
\label{quantity}%
\end{equation}
The proof proceeds as follows.

\begin{itemize}
\item First, from the Schmidt Decomposition Theorem \cite{mosca}, if $\psi$ is
a vector in a tensor product space $\mathcal{H}_{1}\otimes\mathcal{H}_{2}$,
then there exists an orthonormal basis $\left\{  \varphi_{i}\right\}  $ with
$1\leq i\leq n_{\mathcal{H}_{1}}$ for $\mathcal{H}_{1}$ where $n_{\mathcal{H}%
_{1}}\overset{\text{def}}{=}\dim_{%
\mathbb{C}
}\mathcal{H}_{1}$, and an orthonormal basis $\left\{  \theta_{j}\right\}  $
with $1\leq j\leq n_{\mathcal{H}_{2}}$ for $\mathcal{H}_{2}$ where
$n_{\mathcal{H}_{2}}\overset{\text{def}}{=}\dim_{%
\mathbb{C}
}\mathcal{H}_{2}$, and non-negative \emph{real} numbers $\left\{
c_{i}\right\}  $ so that%
\begin{equation}
\psi=\sum_{k}c_{k}\varphi_{k}\otimes\theta_{k}\text{.} \label{state}%
\end{equation}
The coefficients $c_{k}$ are called Schmidt coefficients, and the number of
terms in the sum in Eq. (\ref{state}) will be at most the minimum of
$n_{\mathcal{H}_{1}}$ and $n_{\mathcal{H}_{2}}$. Without loss of generality,
in order to have a non-product state of two-particle systems, assume that at
least two coefficients are nonzero, $c_{1}\neq0\neq c_{2}$ while $c_{k}=0$ for
any $k>2$. We point out that, in principle, coefficients $c_{k}$ with $k>2$
could also be nonzero. However, since the original Clauser-Horne-Shimony-Holt
(CHSH, \cite{chsh}) version of Bell's inequality can be violated in any two
dimensional Hilbert subspace with nonzero Schmidt coefficient \cite{bell64,
prlgisin}, without loss of generality, one can restrict the discussion to the
case in which $c_{k}=0$ for any $k>2$. For the sake of completeness, we point
out that the projection of the joint state of n pairs of particles onto a
subspace spanned by states having a common Schmidt coefficients is a key step
in the so-called Schmidt projection method, a technique used to study
entanglement concentration in any pure state of a bipartite system
\cite{fuck01}. Furthermore, for a practical demonstration of the violation of
the CHSH version of Bell's inequality in $2D$ subspaces of an
higher-dimensional orbital angular momentum Hilbert space, we refer to
\cite{fuck02}.

\item Second, recall that for any separable quantum state, the measure of
entanglement is zero. Furthermore, the behavior of entanglement remains
unchanged under simple local transformations, i.e. local unitary
transformations \cite{vedral}. A local unitary transformation simply
represents a change of basis in which we consider the given entangled state. A
change of basis should not change the amount of entanglement that is
accessible to us, because at any given time we could just reverse the basis
change. Therefore in both bases the entanglement should be the same. Given
these remarks, for the sake of convenience and without changing the
entanglement behavior of the state $\psi$, let us apply a local (product)
unitary transformation $U_{1}\otimes U_{2}$ on $\psi$ such that,%
\begin{equation}
\psi\rightarrow\left(  U_{1}\otimes U_{2}\right)  \psi=\left(  U_{1}\otimes
U_{2}\right)  \left(  c_{1}\varphi_{1}\otimes\theta_{1}+c_{2}\varphi
_{2}\otimes\theta_{2}\right)  \text{,}%
\end{equation}
where,%
\begin{equation}
\left(  U_{1}\otimes U_{2}\right)  \left(  c_{1}\varphi_{1}\otimes\theta
_{1}+c_{2}\varphi_{2}\otimes\theta_{2}\right)  =c_{1}U_{1}\varphi_{1}\otimes
U_{2}\theta_{1}+c_{2}U_{1}\varphi_{2}\otimes U_{2}\theta_{2}=c_{1}\left\vert
+-\right\rangle +c_{2}\left\vert -+\right\rangle \text{.} \label{sopra}%
\end{equation}
In the particular case of spin $1/2$ systems, we have that $\left\vert
+\right\rangle $ and $\left\vert -\right\rangle $ in\ Eq. (\ref{sopra}) are
given by,
\begin{equation}
\left\vert +\right\rangle \overset{\text{def}}{=}\binom{1}{0}\text{, and
}\left\vert -\right\rangle \overset{\text{def}}{=}\binom{0}{1}\text{,}%
\end{equation}
respectively.

\item Third, an arbitrary density matrix $\rho$ for a mixed state qubit can be
written as \cite{nielsen2000, cafaro2012},%
\begin{equation}
\rho\overset{\text{def}}{=}\frac{I+\vec{P}\cdot\vec{\sigma}}{2}\text{,}%
\end{equation}
where $\vec{P}\in%
\mathbb{R}
^{3}$ with $\left\Vert \vec{P}\right\Vert \leq1$ is the Bloch vector for the
state $\rho$, $I$ is the $2\times2$ identity matrix and $\vec{\sigma}$ is the
Pauli matrix vector given by \cite{nielsen2000, cafaro2010, cafaro2014},%
\begin{equation}
\vec{\sigma}=\left[  \sigma_{x}\text{, }\sigma_{y}\text{, }\sigma_{z}\right]
\overset{\text{def}}{\mathbf{=}}\left[  \left(
\begin{array}
[c]{cc}%
0 & 1\\
1 & 0
\end{array}
\right)  \text{, }\left(
\begin{array}
[c]{cc}%
0 & -i\\
i & 0
\end{array}
\right)  \text{, }\left(
\begin{array}
[c]{cc}%
1 & 0\\
0 & -1
\end{array}
\right)  \right]  \text{.} \label{pauli}%
\end{equation}
A state $\rho$ is pure if and only if $\left\Vert \vec{P}\right\Vert =1$. A
density matrix for a pure state can be described as $\rho=\left\vert
\psi\right\rangle \left\langle \psi\right\vert $ where $\left\vert
\psi\right\rangle \left\langle \psi\right\vert $ is an orthogonal projector
since $\rho^{2}=\rho$ and $\rho^{\dagger}=\rho$ (the symbol $\dagger$ denotes
the usual Hermitian conjugation operation in quantum mechanics).
\end{itemize}

Having considered the above mentioned three points, the idea is to check the
correctness of the inequality in Eq. (\ref{bell}) for entangled states $\psi$
in Eq. (\ref{state}) and for a suitable choice of projectors $a$, $a^{\prime}%
$, $b$, and $b^{\prime}$. Consider projectors $a$, $a^{\prime}$, $b$,
$b^{\prime}$ defined as,%
\begin{equation}
a\overset{\text{def}}{=}\frac{1+\vec{a}\cdot\vec{\sigma}}{2}\text{, }%
a^{\prime}\overset{\text{def}}{=}\frac{1+\vec{a}^{\prime}\cdot\vec{\sigma}}%
{2}\text{, }b\overset{\text{def}}{=}\frac{1+\vec{b}\cdot\vec{\sigma}}%
{2}\text{, }b^{\prime}\overset{\text{def}}{=}\frac{1+\vec{b}^{\prime}\cdot
\vec{\sigma}}{2}\text{,} \label{projectors}%
\end{equation}
with $\vec{a}$, $\vec{a}^{\prime}$, $\vec{b}$, $\vec{b}^{\prime}\in%
\mathbb{R}
^{3}$, and
\begin{equation}
\left\Vert \vec{a}\right\Vert =\left\Vert \vec{a}^{\prime}\right\Vert
=\left\Vert \vec{b}\right\Vert =\left\Vert \vec{b}^{\prime}\right\Vert
=1\text{,}%
\end{equation}
so that $a^{2}=a$, $a^{\prime2}=a^{\prime}$, $b^{2}=b$, $b^{\prime2}%
=b^{\prime}$. From Eq. (\ref{projectors}), $P\left(  a\text{, }b\right)  $ in
Eq. (\ref{quantity}) becomes%
\begin{equation}
P\left(  a\text{, }b\right)  \overset{\text{def}}{=}\left\langle \left(
2a-1\right)  \otimes\left(  2b-1\right)  \right\rangle _{\psi}=\left\langle
\left(  \vec{a}\cdot\vec{\sigma}\right)  \otimes\left(  \vec{b}\cdot
\vec{\sigma}\right)  \right\rangle _{\psi}\text{,}%
\end{equation}
with $\left\vert \psi\right\rangle \overset{\text{def}}{=}c_{1}\left\vert
+-\right\rangle +c_{2}\left\vert -+\right\rangle $ and $c_{1}$, $c_{2}\in%
\mathbb{R}
\backslash\left\{  0\right\}  $. Let us compute the explicit expression of
$P\left(  a\text{, }b\right)  $. Using Eq. (\ref{pauli}), we\textbf{ }obtain%
\begin{align}
\left(  \vec{a}\cdot\vec{\sigma}\right)  \otimes\left(  \vec{b}\cdot
\vec{\sigma}\right)   &  =\left(
\begin{array}
[c]{cc}%
a_{z} & a_{x}-ia_{y}\\
a_{x}+ia_{y} & -a_{z}%
\end{array}
\right)  \otimes\left(
\begin{array}
[c]{cc}%
\ b_{z} & b_{x}-ib_{y}\\
b_{x}+ib_{y} & -b_{z}%
\end{array}
\right) \nonumber\\
& \nonumber\\
&  =\left(
\begin{array}
[c]{cccc}%
a_{z}b_{z} & a_{z}\left(  b_{x}-ib_{y}\right)  & b_{z}\left(  a_{x}%
-ia_{y}\right)  & \left(  a_{x}-ia_{y}\right)  \left(  b_{x}-ib_{y}\right) \\
a_{z}\left(  b_{x}+ib_{y}\right)  & -a_{z}b_{z} & \left(  a_{x}-ia_{y}\right)
\left(  b_{x}+ib_{y}\right)  & -b_{z}\left(  a_{x}-ia_{y}\right) \\
b_{z}\left(  a_{x}+ia_{y}\right)  & \left(  a_{x}+ia_{y}\right)  \left(
b_{x}-ib_{y}\right)  & -a_{z}b_{z} & -a_{z}\left(  b_{x}-ib_{y}\right) \\
\left(  a_{x}+ia_{y}\right)  \left(  b_{x}+ib_{y}\right)  & -b_{z}\left(
a_{x}+ia_{y}\right)  & -a_{z}\left(  b_{x}+ib_{y}\right)  & a_{z}b_{z}%
\end{array}
\right)  \text{.}%
\end{align}
Therefore the quantity $P\left(  a\text{, }b\right)  $ becomes,%
\begin{align}
P\left(  a\text{, }b\right)   &  =\left\langle \left(  \vec{a}\cdot\vec
{\sigma}\right)  \otimes\left(  \vec{b}\cdot\vec{\sigma}\right)  \right\rangle
_{\psi}=\left\langle \psi\left\vert \left(  \vec{a}\cdot\vec{\sigma}\right)
\otimes\left(  \vec{b}\cdot\vec{\sigma}\right)  \right\vert \psi\right\rangle
\nonumber\\
& \nonumber\\
&  =\left(
\begin{array}
[c]{cccc}%
0 & c_{1} & c_{2} & 0
\end{array}
\right)  \cdot\nonumber\\
& \nonumber\\
&  \cdot\left(
\begin{array}
[c]{cccc}%
a_{z}b_{z} & a_{z}\left(  b_{x}-ib_{y}\right)  & b_{z}\left(  a_{x}%
-ia_{y}\right)  & \left(  a_{x}-ia_{y}\right)  \left(  b_{x}-ib_{y}\right) \\
a_{z}\left(  b_{x}+ib_{y}\right)  & -a_{z}b_{z} & \left(  a_{x}-ia_{y}\right)
\left(  b_{x}+ib_{y}\right)  & -b_{z}\left(  a_{x}-ia_{y}\right) \\
b_{z}\left(  a_{x}+ia_{y}\right)  & \left(  a_{x}+ia_{y}\right)  \left(
b_{x}-ib_{y}\right)  & -a_{z}b_{z} & -a_{z}\left(  b_{x}-ib_{y}\right) \\
\left(  a_{x}+ia_{y}\right)  \left(  b_{x}+ib_{y}\right)  & -b_{z}\left(
a_{x}+ia_{y}\right)  & -a_{z}\left(  b_{x}+ib_{y}\right)  & a_{z}b_{z}%
\end{array}
\right)  \cdot\nonumber\\
& \nonumber\\
&  \cdot\left(
\begin{array}
[c]{c}%
0\\
c_{1}\\
c_{2}\\
0
\end{array}
\right) \nonumber\\
& \nonumber\\
&  =-\left(  c_{1}^{2}+c_{2}^{2}\right)  a_{z}b_{z}+c_{1}c_{2}\left(
2a_{x}b_{x}+2a_{y}b_{y}\right)  \text{,}%
\end{align}
that is,%
\begin{equation}
P\left(  a\text{, }b\right)  =-\left(  c_{1}^{2}+c_{2}^{2}\right)  a_{z}%
b_{z}+c_{1}c_{2}\left(  2a_{x}b_{x}+2a_{y}b_{y}\right)  \text{.}%
\end{equation}
Normalization of the wave function demands $\left\langle \psi\left\vert
\psi\right.  \right\rangle =1$. Therefore, we have $c_{1}^{2}+c_{2}^{2}=1$
and,%
\begin{equation}
P\left(  a\text{, }b\right)  =+2c_{1}c_{2}\left(  a_{x}b_{x}+a_{y}%
b_{y}\right)  -a_{z}b_{z}\text{.} \label{first correction}%
\end{equation}
This simple mathematical computation leading to Eq. (\ref{first correction})
leads to rectify the incorrect sign that appears in \cite{gisin91}. Following
Gisin, we assume that convenient expressions for the Bloch vectors are given
by,%
\begin{align}
\vec{a}  &  =\left(  a_{x}\text{, }a_{y}\text{, }a_{z}\right)  =\left(
\sin\alpha\text{, }0\text{, }\cos\alpha\right)  \text{,}\nonumber\\
\vec{b}  &  =\left(  b_{x}\text{, }b_{y}\text{, }b_{z}\right)  =\left(
\sin\beta\text{, }0\text{, }\cos\beta\right)  \text{,} \label{vector}%
\end{align}
with $\alpha=0$, $\alpha^{\prime}=\pm\pi/2$ where the sign is the \emph{same}
as that of $c_{1}c_{2}$ (note that due to the previous sign mistake, in
\cite{gisin91} it is reported that $\alpha^{\prime}=\pm\pi/2$ where the sign
is the \emph{opposite} of that of $c_{1}c_{2}$). For possible alternative
parametrizations of the Bloch vectors, we refer to Appendix A. Let us now
consider the expression given by $\left\vert P\left(  a\text{, }b\right)
-P\left(  a\text{, }b^{\prime}\right)  \right\vert +P\left(  a^{\prime}\text{,
}b\right)  +P\left(  a^{\prime}\text{, }b^{\prime}\right)  $. Using Eqs.
(\ref{quantity}), (\ref{projectors}), and (\ref{vector}), we obtain%
\begin{equation}
\left\vert P\left(  a\text{, }b\right)  -P\left(  a\text{, }b^{\prime}\right)
\right\vert =\left\vert -\cos\beta+\cos\beta^{\prime}\right\vert =\left\vert
-\left(  \cos\beta-\cos\beta^{\prime}\right)  \right\vert =\left\vert
\cos\beta-\cos\beta^{\prime}\right\vert \text{.}%
\end{equation}
Furthermore, using the same line of reasoning, $P\left(  a^{\prime}\text{,
}b\right)  +P\left(  a^{\prime}\text{, }b^{\prime}\right)  $ becomes%
\begin{align}
P\left(  a^{\prime}\text{, }b\right)  +P\left(  a^{\prime}\text{, }b^{\prime
}\right)   &  =\left(  2c_{1}c_{2}\sin\alpha^{\prime}\sin\beta-\cos
\alpha^{\prime}\cos\beta\right)  +\left(  2c_{1}c_{2}\sin\alpha^{\prime}%
\sin\beta^{\prime}-\cos\alpha^{\prime}\cos\beta^{\prime}\right) \nonumber\\
& \nonumber\\
&  =2c_{1}c_{2}\sin\alpha^{\prime}\left(  \sin\beta+\sin\beta^{\prime}\right)
-\cos\alpha^{\prime}\left(  \cos\beta+\cos\beta^{\prime}\right)  \text{,}%
\end{align}
that is,%
\begin{equation}
P\left(  a^{\prime}\text{, }b\right)  +P\left(  a^{\prime}\text{, }b^{\prime
}\right)  =2c_{1}c_{2}\sin\alpha^{\prime}\left(  \sin\beta+\sin\beta^{\prime
}\right)  -\cos\alpha^{\prime}\left(  \cos\beta+\cos\beta^{\prime}\right)
\text{.}%
\end{equation}
Assuming $c_{1}c_{2}>0$ and $\alpha^{\prime}=+\pi/2$, yields%
\begin{equation}
P\left(  a^{\prime}\text{, }b\right)  +P\left(  a^{\prime}\text{, }b^{\prime
}\right)  =2\left\vert c_{1}c_{2}\right\vert \left(  \sin\beta+\sin
\beta^{\prime}\right)  \text{.}%
\end{equation}
Moreover, assuming $c_{1}c_{2}<0$ and $\alpha^{\prime}=-\pi/2$, we get%
\begin{equation}
P\left(  a^{\prime}\text{, }b\right)  +P\left(  a^{\prime}\text{, }b^{\prime
}\right)  =-2c_{1}c_{2}\left(  \sin\beta+\sin\beta^{\prime}\right)  \text{,}%
\end{equation}
with $-2c_{1}c_{2}>0$. Therefore $2\left\vert c_{1}c_{2}\right\vert
=-2c_{1}c_{2}>0$, and%
\begin{equation}
P\left(  a^{\prime}\text{, }b\right)  +P\left(  a^{\prime}\text{, }b^{\prime
}\right)  =2\left\vert c_{1}c_{2}\right\vert \left(  \sin\beta+\sin
\beta^{\prime}\right)  \text{.}%
\end{equation}
In summary, the expression for $\left\vert P\left(  a\text{, }b\right)
-P\left(  a\text{, }b^{\prime}\right)  \right\vert +P\left(  a^{\prime}\text{,
}b\right)  +P\left(  a^{\prime}\text{, }b^{\prime}\right)  $ is given by
\begin{equation}
\left\vert P\left(  a\text{, }b\right)  -P\left(  a\text{, }b^{\prime}\right)
\right\vert +P\left(  a^{\prime}\text{, }b\right)  +P\left(  a^{\prime}\text{,
}b^{\prime}\right)  =\left\vert \cos\beta-\cos\beta^{\prime}\right\vert
+2\left\vert c_{1}c_{2}\right\vert \left(  \sin\beta+\sin\beta^{\prime
}\right)  \text{.}%
\end{equation}
At this point, choose $\beta$ and $\beta^{\prime}$ such that $\sin\beta>0$ and
$\sin\beta^{\prime}>0$ so that $2\left\vert c_{1}c_{2}\right\vert \left(
\sin\beta+\sin\beta^{\prime}\right)  >0$. Take also $-\cos\beta^{\prime}%
=\cos\beta=\left(  1+4\left\vert c_{1}c_{2}\right\vert \right)  ^{-1/2}$. We
finally obtain,%
\begin{equation}
\left\vert P\left(  a\text{, }b\right)  -P\left(  a\text{, }b^{\prime}\right)
\right\vert +P\left(  a^{\prime}\text{, }b\right)  +P\left(  a^{\prime}\text{,
}b^{\prime}\right)  \geq2\left(  1+4\left\vert c_{1}c_{2}\right\vert \right)
^{-1/2}>2\text{.}%
\end{equation}
This concludes the proof.

\section{Conclusions}

In this article, we presented a simple and explicit reexamination of Gisin's
1991 original proof concerning the violation of Bell's inequality for any
entangled state of two-particle systems. Gisin's original work on Bell's
inequality was presented in a very synthetic and intuitive manner. After so
many years, Gisin's work continues to be highly regarded also in pure research
\cite{rudolph}. For this reason, we have considered it especially helpful for
didactic purposes. For a different alternative pedagogical presentation of
Bell's inequality, we refer to \cite{selleri, maccone13}. We remark that we
have simply expanded the original proof and in doing so, we have rectified
with Eq. (\ref{first correction}) an incorrect mathematical sign in the
expression of $P\left(  a\text{, }b\right)  $ in \cite{gisin91} that now
becomes,%
\begin{equation}
P\left(  a\text{, }b\right)  =+2c_{1}c_{2}\left(  a_{x}b_{x}+a_{y}%
b_{y}\right)  -a_{z}b_{z}\text{.} \label{pab}%
\end{equation}
The incorrectness of the sign in $P\left(  a\text{, }b\right)  $ as reported
in \cite{gisin91} generates and propagates an incorrect statement. As a
consequence of Eq. (\ref{pab}), we have also corrected such a subsequent
statement. Namely, the correct statement becomes%
\begin{equation}
\alpha^{\prime}=\pm\pi/2\text{ where the sign is the \emph{same }as that of
}c_{1}c_{2}\text{.} \label{fu}%
\end{equation}
Notwithstanding these two corrections presented in Eqs. (\ref{pab}) and
(\ref{fu}), the physical conclusions provided by Gisin remain unaltered.

\begin{acknowledgments}
C. Cafaro thanks Nicolas Gisin for his kindness in providing useful
correspondence and for pointing out reference \cite{gisin92}. C. Cafaro is
also grateful to Ariel Caticha for helpful discussions.
\end{acknowledgments}

\bigskip\pagebreak

\appendix

\section{Alternative parametrizations of the Bloch vectors}

In general, the Bloch vectors $\vec{a}$ and $\vec{b}$ in $%
\mathbb{R}
^{3}$ that appear in Eq. (\ref{projectors}) can be parametrized in terms of
the polar angle $\theta$ with $0\leq\theta\leq\pi$ and the azimuthal angle
$\varphi$ with $0\leq\varphi<2\pi$ as follows,%
\begin{equation}
\vec{a}=\left(  a_{x}\text{, }a_{y}\text{, }a_{z}\right)  =\left(  \sin
\theta_{1}\cos\varphi_{1}\text{, }\sin\theta_{1}\sin\varphi_{1}\text{, }%
\cos\theta_{1}\text{ }\right)  \text{,} \label{aa1}%
\end{equation}
and,%
\begin{equation}
\vec{b}=\left(  b_{x}\text{, }b_{y}\text{, }b_{z}\right)  =\left(  \sin
\theta_{2}\cos\varphi_{2}\text{, }\sin\theta_{2}\sin\varphi_{2}\text{, }%
\cos\theta_{2}\text{ }\right)  \text{,} \label{aa2}%
\end{equation}
respectively. Using Eqs. (\ref{aa1}) and (\ref{aa2}), the quantity $P\left(
a\text{, }b\right)  $ that appears in Eq. (\ref{bell}) becomes%
\begin{equation}
P\left(  a\text{, }b\right)  =2c_{1}c_{2}\left[  \sin\theta_{1}\sin\theta
_{2}\cos\varphi_{1}\cos\varphi_{2}+\sin\theta_{1}\sin\theta_{2}\sin\varphi
_{1}\sin\varphi_{2}\right]  -\cos\theta_{1}\cos\theta_{2}\text{.}%
\end{equation}
Similar expressions are valid for $P\left(  a\text{, }b^{\prime}\right)  $,
$P\left(  a^{\prime}\text{, }b\right)  $, and $P\left(  a^{\prime}\text{,
}b^{\prime}\right)  $ in Eq. (\ref{bell}). It would be interesting to provide
a thorough numerical investigation for the conditions on the vectors $\vec{a}$
and $\vec{b}$ in Eqs. (\ref{aa1}) and (\ref{aa2}) such that Bell's inequality
is violated. Here, however, we simply point out that Gisin's original
parametrization is not unique and alternative parametrizations may be
considered, preserving essentially Gisin's original line of reasoning. Gisin's
choice was to set $\varphi_{1}=\varphi_{2}=0$ in Eqs. (\ref{aa1}) and
(\ref{aa2}), respectively, leading to%
\begin{equation}
\vec{a}=\left(  \sin\theta_{1}\text{, }0\text{, }\cos\theta_{1}\text{
}\right)  \text{, and }\vec{b}=\left(  \sin\theta_{2}\text{, }0\text{, }%
\cos\theta_{2}\text{ }\right)  \text{.}%
\end{equation}
Two natural alternatives are the following: i) $\varphi_{1}=\varphi_{2}=\pi
/2$; ii) $\theta_{1}=\theta_{2}=\pi/2$. In the former and latter cases, we
obtain%
\begin{equation}
\vec{a}=\left(  0\text{, }\sin\theta_{1}\text{, }\cos\theta_{1}\text{
}\right)  \text{, and }\vec{b}=\left(  0\text{, }\sin\theta_{2}\text{, }%
\cos\theta_{2}\text{ }\right)  \text{,}%
\end{equation}
and,%
\begin{equation}
\vec{a}=\left(  \cos\varphi_{1}\text{, }\sin\varphi_{1}\text{, }0\text{
}\right)  \text{, and }\vec{b}=\left(  \cos\varphi_{2}\text{, }\sin\varphi
_{2}\text{, }0\text{ }\right)  \text{,}%
\end{equation}
respectively. Following the line of reasoning presented in this manuscript, it
is relatively straightforward to show that Gisin's argument concerning Bell's
inequality applies essentially in a similar manner to these alternative
parametrizations of the Bloch vectors.

\bigskip

\bigskip

\bigskip

\bigskip

\bigskip

\bigskip


\bigskip

\begin{thebibliography}{99}                                                                                               %


\bibitem {peres2004}A. Peres and D. R. Terno, \emph{Quantum information and
relativity theory}, Rev. Mod. Phys. \textbf{76}, 93 (2004).

\bibitem {sakurai}J. J. Sakurai, \emph{Modern Quantum Mechanics}, The
Benjamin/Cummings Publishing Company, Inc. (1985).

\bibitem {colella1975}R. Colella, A. W. Overhauser, and S. A. Werner,
\emph{Observation of gravitationally induced quantum interference}, Phys. Rev.
Lett. \textbf{34}, 1472 (1975).

\bibitem {brukner14}C. Brukner, \emph{Quantum causality}, Nature Physics
\textbf{10}, 259 (2014).

\bibitem {bohr35}N. Bohr, \emph{Can quantum-mechanical description of physical
reality be considered complete?}, Phys. Rev. \textbf{48}, 696 (1935).

\bibitem {bohr37}N. Bohr, \emph{Causality and complementarity}, Philosophy of
Science \textbf{4}, 289 (1937).

\bibitem {bohr50}N. Bohr, \emph{On the notions of causality and
complementarity}, Science \textbf{111}, 51 (1950).

\bibitem {wheeler63}J. A. Wheeler, \emph{No fugitive and cloistered virtue- A
tribute to Niels Bohr}, Physics Today \textbf{16}, 30 (1963).

\bibitem {vaccaro10}J. A. Vaccaro, \emph{Group theoretic formulation of
complementarity}, arXiv:quant-ph/1012.3532 (2010).

\bibitem {epr}A. Einstein, B. Podolsky, and N. Rosen, \emph{Can
quantum-mechanical description of physical reality be considered complete?},
Phys. Rev. \textbf{47}, 777 (1935).

\bibitem {bell64}J. S. Bell, \emph{On the Einstein Podolsky Rosen paradox},
Physics \textbf{1}, 195 (1964).

\bibitem {gisin91}N. Gisin, \emph{Bell's inequality holds for all non-product
states}, Phys. Lett. \textbf{A154}, 201 (1991).

\bibitem {peres95}A. Peres, \emph{Quantum Theory: Concepts and Methods},
Kluwer Academic Publishers (1995).

\bibitem {peres04}A. Peres, \emph{Quantum information and general relativity},
arXiv:quant-ph/0405127 (2004).

\bibitem {stapp75}H. P. Stapp, \emph{Bell's theorem and world process}, Il
Nuovo Cimento \textbf{29}, 270 (1975).

\bibitem {gisin92}N. Gisin and A. Peres, \emph{Maximal violation of Bell's
inequality for arbitrary large spin}, Phys. Lett. \textbf{A162}, 15 (1992).

\bibitem {mosca}P. Kaye, R. Laflamme, and M. Mosca, \emph{An Introduction to
Quantum Computing}, Oxford University Press (2007).

\bibitem {chsh}J. F. Clauser, M. A. Horne, A. Shimony, and R. A. Holt,
\emph{Proposed experiment to test local hidden-variable theories}, Phys. Rev.
Lett. \textbf{23}, 880 (1969).

\bibitem {prlgisin}D. Collins, N. Gisin, N. Linden, S. Massar, and S. Popescu,
\emph{Bell inequalities for arbitrarily high-dimensional systems}, Phys. Rev.
Lett. \textbf{88}, 040404 (2002).

\bibitem {fuck01}C. H. Bennett, H. J. Bernstein, S. Popescu and B. Schumacher,
\emph{Concentrating partial entanglement by local operations}, Phys. Rev.
\textbf{A53, }2046 (1996).

\bibitem {fuck02}J. Leach, B. Jack, J. Romero, M. Ritsch-Marte, R. W. Boyd, A.
K. Jha, S. M. Barnett, S. Franke-Arnold and M. J. Padgett, \emph{Violation of
a Bell inequality in two-dimensional orbital angular momentum state-spaces},
Opt. Express \textbf{17, }8287 (2009).

\bibitem {vedral}J. Dunningham and V. Vedral, \emph{Introductory Quantum
Physics and Relativity}, Imperial College Press (2011).

\bibitem {nielsen2000}M. A. Nielsen and I. L. Chuang, \emph{Quantum
Computation and Quantum Information}, Cambridge University Press (2000).

\bibitem {cafaro2012}C. Cafaro and S. Mancini, \emph{Characterizing the
depolarizing quantum channel in terms of Riemannian geometry}, Int. J. Geom.
Methods Mod. Phys. \textbf{9}, 1260020 (2012).

\bibitem {cafaro2010}C. Cafaro\textbf{ }and S. Mancini, \emph{Quantum
stabilizer codes for correlated and asymmetric depolarizing errors}, Phys.
Rev. \textbf{A82}, 012306 (2010).

\bibitem {cafaro2014}C. Cafaro and P. van Loock, \emph{Approximate quantum
error correction for generalized amplitude-damping errors}, Phys. Rev.
\textbf{A89}, 022316 (2014).

\bibitem {rudolph}S. Jevtic and T. Rudolph, \emph{How Einstein and/or
Schrodinger should have discovered Bell's Theorem in 1936}, JOSA \textbf{B32},
A50 (2015).

\bibitem {selleri}V. Capasso, D. Fortunato, and F. Selleri, \emph{Sensitive
observables of quantum mechanics}, Int. J. Theor. Phys. \textbf{7}, 319 (1973).

\bibitem {maccone13}L. Maccone, \emph{A simple proof of Bell's inequality},
Am. J. Phys. \textbf{81}, 854 (2013).
\end{thebibliography}
\end{document}